\documentclass[traditabstract]{aa} 
\usepackage{txfonts}
\usepackage{graphicx}
\newcommand{\subr}{erg s$^{-1}$ Mpc$^{-2}$ }
\newcommand{\ergs}{erg s$^{-1}$ }

\begin{document}

\title{
Observed abundance of X-ray low surface brightness clusters in optical, X-ray, and SZ selected
samples
}
\titlerunning{Completeness of SZ and X-ray surveys}
\author{S. Andreon$^{1}$\thanks{E-mail:stefano.andreon@inaf.it}, G. Trinchieri$^{1}$, A. Moretti$^{1}$}
\authorrunning{S. Andreon et al.}
\institute{
$^1$ INAF--Osservatorio Astronomico di Brera, via Brera 28, 20121, Milano, Italy\\
}
\date{Accepted ... Received ..; in original form ..}
\abstract{
The comparison of the properties of galaxy cluster samples selected using observations in different wavebands
may shed light on potential biases of the way in which the samples are assembled. For this comparison, we introduce a new observable 
that does not require previous knowledge of the cluster mass: the X-ray
mean surface brightness within the central 300 kpc.
We found that clusters with low surface brightness, defined as those with a mean surface brightness below 43.35 \subr ,
are  about one quarter of the whole cluster population in a sample of 32 clusters in the nearby Universe, selected independently of the
intracluster medium properties. Almost no example of a low central surface brightness cluster exists instead in
two X-ray selected samples, one sample based on XMM-Newton XXL-100 survey data and the other on full-depth eROSITA eFEDS data, although 
these clusters are known to exist in the range of redshift and mass as probed by these two surveys.
Furthermore,  the
Sunayev-Zeldovich Atacama Cosmology Telescope cluster survey 
is even more selective than the previous two samples because it  does not even include clusters with
intermediate surface brightness, which are instead present in X-ray selected samples that explore the same volume of the Universe.
Finally, a measure of the
mean surface brightness, which is obtained without knowledge of the mass, 
proves to be effective in
narrowing the number of clusters to be followed-up because it recognizes those
with a low gas fraction or with a low X-ray luminosity for their mass. Identifying these would otherwise require knowledge of the mass for all clusters.
}
\maketitle

\section{Introduction}

Our current knowledge of the properties of the intracluster medium of galaxy clusters comes primarily from detailed
studies of clusters selected through the intracluster medium, either in
emission (in X-ray) or via the effect the intracluster medium has
on photons of the cosmic microwave background (Sunayev-Zeldovich effect; SZ hereafter, Sunyaev \& Zeldovich 1972). 
It is now generally appreciated 
 that clusters selected in  
X-ray surveys offer a biased view of the cluster population, although
considerable effort has been made to take into account and correct for the inherent biases
(Pacaud et al. 2007;  Stanek et al. 2006;   
Andreon, Trinchieri \& Pizzolato 2011; Andreon \& Moretti 2011; 
Eckert et al. 2011; Planck Collaboration 2011, 2012;
Maughan et al. 2012; Anderson et al. 2015; Andreon et al. 2016). 
The bias occurs
because at a given mass, brighter-than-average clusters
are easier to select and be made part of a sample, 
while fainter-than-average clusters are easily missed. 
It is difficult to correct for the bias because the correction depends on assumptions about 
the unseen population, namely the clusters that are faint for their mass  
(Vikhlinin et al. 2009; Andreon et al. 2011, 2016, 2017). Andreon et al. (2022) showed
that the covariance between
detectability and location of a cluster in the mass-temperature diagram is strong, whereas 
the luminosity-temperature plane
is largely unaffected by the missed population. This
emphasizes that the missing population 
behaves differently in the various scaling relations. Using an
X-ray unbiased sample (XUCS hereafter) selected from velocity dispersion measurements, we unveiled an even larger variety
of clusters at a given mass (Andreon et al. 2016, 2017, hereafter Paper I and Paper II) 
that was lost in previous surveys because their surface brightness is low. 
Paper II showed these differences to be related to
the gas fraction: Clusters with a low surface brightness also have a low gas fraction.
Some of these low surface brightness clusters, with some new additions, are being rediscovered 
by other authors from independent data (Lietzen et al. 2024). Low surface objects of lower mass
are also being discovered (O'Sullivan et al. 2017, Pearson et al. 2017, Xu et al. 2018, 2022, Capasso et al. 2020; Crossett et al. 2022).

Cluster samples selected using the SZ effect
 are said to offer a less biased view because the SZ signal in simulations is tightly correlated to the mass
(e.g., Motl et al. 2005; Nagai et al. 2006; Angulo et al. 2009). They show a larger
variety in X-ray luminosity at a fixed mass
than X-ray selected samples  (e.g., Planck Collaboration 2011, 2012), but as shown by recent analyses (e.g., Andreon et al. 2016, 2017; Xu et al. 2018, 2022; Orlowski-Scherer et al. 2021) and in this work, even the SZ-selection fails to sample the full range. 
While much of the literature identified integrated pressure with mass,
the existence of massive clusters that are undetectable in current SZ surveys, that is, which have a low brightness in their SZ signal, is slowly starting to be allowed in collective analyses of SZ samples (Orlowski-Scherer et al. 2021, Grandis et al. 2021). 

We extensively studied one cluster with a low surface brightness, CL2015 (Andreon et al. 2019, hereafter Paper IV). 
Its core-excised X-ray luminosity is low for its mass $M_{500}$, $12\sigma$ below the mean relation derived from 
the X-ray selected sample of Pratt et al. (2009), but only $1 \sigma$ 
below that derived for the X-ray unbiased sample. CL2015 differs from  X-ray selected clusters 
in two aspects: First, the total mass profile has a very low concentration. This feature is shared by weak-lensing 
selected clusters, which have lower concentrations than X-ray selected
clusters on average (Miyazaki et al. 2018). Second, the gas pressure profile and integrated pressure (measured by the
total $Y_{SZ}$) are greatly depressed (by a factor of three). Similar objects were recently reported at
$z=1.75$ (Andreon et al. 2021) and at $z=1.80$ (Andreon et al. 2022), as well as several other less clear examples at $z\approx 1$ (Dicker et al. 2020; Di Mascolo et al. 2020). 
The existence of several clusters with
depressed pressure profiles has profound cosmological and astrophysical
consequences because
an overestimate by 15\% of the average pressure profile is enough to resolve the tension between cosmological
parameters derived from CMB anisotropies and cluster abundances (Ruppin et al. 2019). 
We currently observe only one-third of the clusters expected from CMB cosmology (e.g., Planck collaboration 2014); if a larger number of clusters with a low X-ray surface brightness exist, the tension could be attenuated or resolved.

The comparison of the properties of galaxy cluster samples selected using observations in different wavebands
may shed light on potential biases of the way in which samples are assembled. Several works analyzed the overlap between cluster samples selected at different wavelengths.  
Donahue et al. (2002), Gilbank et al. (2004), Ryfoff et al. (2008), Sadibekova et al. (2014), Willis et al. (2021), 
and Upsdell et al. (2023), among others,  
studied the overlap between optical and X-ray selected cluster samples. However, the X-ray data were shallow,
and this led to poorly constrained results because objects were undetected, regardless of whether the two populations
overlapped or were widely different (see Andreon \& Moretti 2011 for a discussion). Sadibekova et al. (2022) and Upsdell et al. (2023) reported that
the richest optically selected clusters are X-ray detected when they are in a redshift range in which the considered X-ray data are informative. Donahue et al. (2020) and Willis et al. (2021) instead emphasized that the X-ray selection misses some bona fide optically selected and X-ray bright  clusters.
Andreon \& Moretti (2012) used deeper X-ray data for the considered 
clusters and concluded that the optical and X-ray selected populations largely overlap, although with a large (0.51 dex) scatter in X-ray luminosity at a fixed richness, which makes their X-ray detection hard and poorly predictable. Andreon et al. (2016) confirmed the above large variety using targeted X-ray observations of a velocity-dispersion-selected cluster sample, XUCS.
In this work,
we intend to proceed in a similar way to what was adopted for galaxies, that is, we create histograms of the number of objects per unit central surface brightness and learn selection effects from the observed differences in the histograms (e.g., McGaugh 1996, O'Neil, Andreon, \& Cuillandre 2003).
For this comparison, we introduce in Sec.~2 four samples
that were selected differently, and a new observable, the X-ray
mean surface brightness within the central 300 kpc. In Sec.~3 we compare the samples to understand how many clusters with a low surface brightness are present.
In Sec.~4 we discuss whether low surface brightness clusters are
absent in these samples because they are missed or because they do not exist in the redshift and mass ranges sampled by these surveys.  Sec.~4 summarizes the results.
Throughout this paper, we assume $\Omega_M=0.3$, $\Omega_\Lambda=0.7$, and $H_0=70$ km s$^{-1}$ Mpc$^{-1}$.

\section{Samples}

To explore the sensitivity of  optical, X-ray,  and SZ-selected samples to clusters with a low X-ray surface brightness,
we considered  four samples: 
\begin{enumerate}
\item  First of all,  we considered
the XUCS sample. It was selected from 
the Sloan digital spectroscopic survey and consists of 34 clusters in the very
nearby Universe ($0.050 < z < 0.135$), characterized by more  than 50
concordant galaxy redshifts whithin 1 Mpc,  a velocity dispersion of
the members $\sigma_v>500$ km/s (see Paper I and Paper II), and low line-of-sight 
Galactic absorption. 
The probability of a cluster to be part of the sample is independent of
its X-ray luminosity or any X-ray property. 
The X-ray properties were obtained later with targeted observations with the Neil Gehrels Swift Observatory,
except for a handful of clusters that had adequate archival  XMM-Newton or Chandra data. 
In the current paper we use the core radii derived in Paper I. Briefly, we fit
a modified beta model with $\beta=2/3$ to individual photons in the [0.5-2] keV band,
accounting 
for excised regions, gaps, and variation in
the exposure time and background. The modified beta model has a power-law-type 
cusp at the center to allow for a cool core.
Weak priors were taken for all parameters. Two examples of fit of the radial profiles are shown in Figure 4 in Paper I.
The median core radius error is 12\%. To offer a glimpse of the spread of the quality of the determinations, the error interquartile range is (7,15)\%, the worst
determination has an error of 31\%. The values of the core radii of the three clusters  measured from observations with different telescopes are consistent with each other (they differ by 2.5, 1.3, and 0.2 $\sigma$).  
Using the same radial fitting model as adopted
for computing X-ray luminosities, we measured the mean [0.5-2] keV X-ray surface brightness 
within an aperture radius of 300 kpc, $\mu_{300}$. The aperture radius was chosen to be equal to
other literature determinations (e.g., Giles et al. 2016; Liu et al. 2021).
The median $\mu_{300}$ error is 0.03 dex, the error interquartile range is (0.01,0.04) dex, and the worst
determination has an error of 0.09 dex.
Table~\ref{tab1} lists derived core radii and $\mu_{300}$ brightnesses. 
As in Andreon et al. (2022), 
we excluded two clusters from the further analysis: CL1022 because it is bimodal (two X-ray peaks), and CL2081 because it is too faint
to derive a robust estimate of the temperature.

\item
As a first X-ray selected sample, we considered the XXL-100 sample (Pierre et al. 2016), which is formed by
the 100 brightest (in an aperture of 2 arcmin)
clusters in the 50 deg$^2$ of the XXL survey (with $\gtrsim 10$ ks observations with XMM-Newton). 
The luminosities 
were taken from  Giles et al. (2016), 
whereas the core radii come from the XXL database\footnote{http://cosmosdb.iasf-milano.inaf.it/XXL/}, where they are listed without errors. 
These core radii were used by the XXL collaboration, for example, in a
XXL-100 detectability study (Pacaud et al. 2016), and the  luminosities were used in several papers,
including by Giles et al. (2016).

\item 
As a second X-ray selected sample, we considered
the eROSITA\footnote{extended ROentgen Survey with an Imaging Telescope Array} Final Equatorial Depth Survey (eFEDS: Brunner et al. 2021), which covers 
about 140 square degrees and has a sensitivity that exceeds what is expected from the 
the final eROSITA full survey in the equatorial region.
The authors of the eFEDS sample (Liu et al. 2021; Klein et al. 2021)
quoted core radii and X-ray luminosities within 300 kpc.
We only considered $F_{cont}<0.2$ clusters to reduce false detections, as suggested by the authors, 
and a minimum of 200 counts to have reliable estimates of the brightness and
core radius. This sample consists of 85 clusters.

\begin{figure}
\centerline{\includegraphics[width=9truecm]{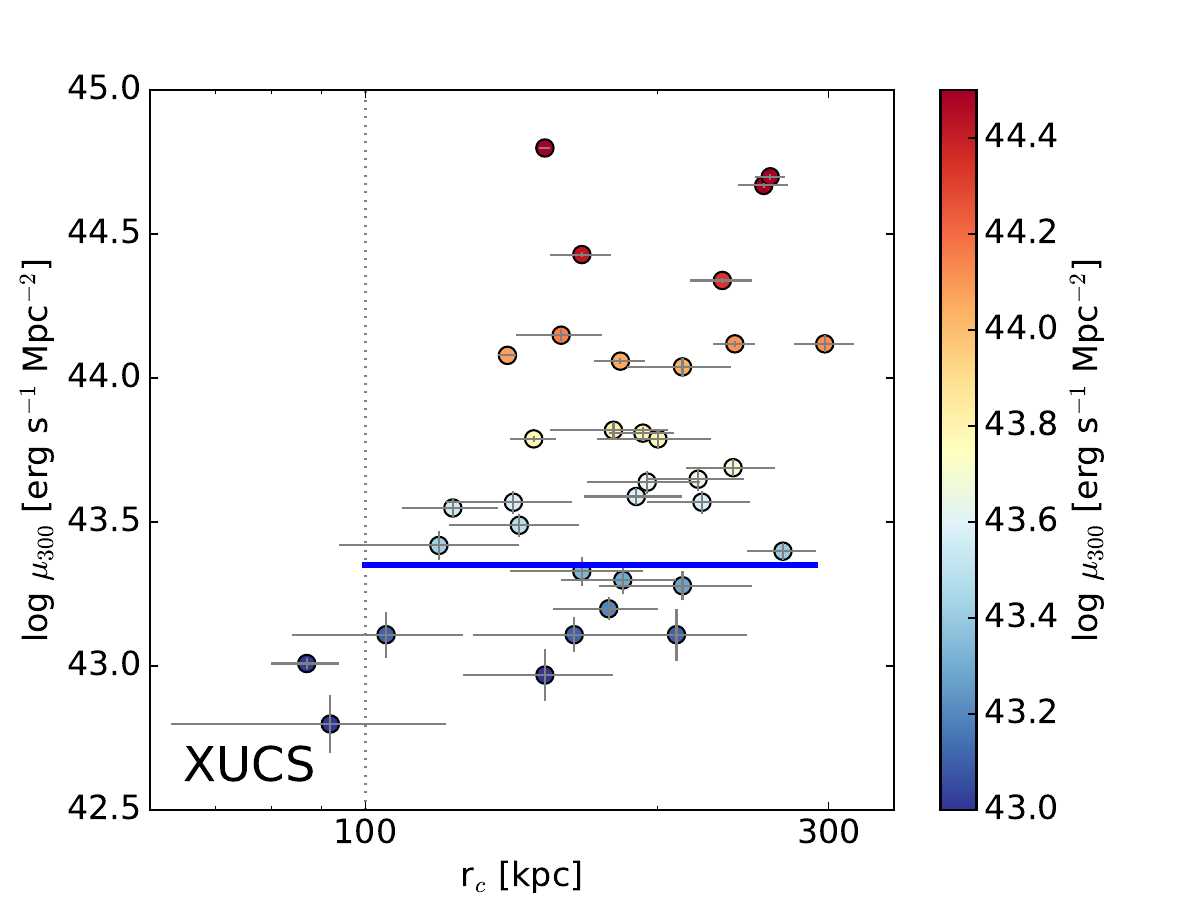}}
\caption[h]{Mean surface brightness vs. core radius. 
The mean surface brightness of about  one quarter of the clusters with $r_c>100$ kpc
is below 43.35 \subr (blue line) in the [0.5-2] keV band within a radius of 300 kpc. The vertical dotted line marks the minimum core radius considered for computing the fraction of low surface brightness clusters.
Three clusters (observed by two telescopes) appear twice in the figure,  but were counted only once in the accounting of the low surface brightness fraction. The individual values can be found in Table 1.  The points are color-coded by $\mu_{300}$ to facilitate the comparison
with Fig.~\ref{fig:LxMoffcod}.
}
\label{fig:murcmucod}
\end{figure}

\item  To test the
SZ selection, which 
has been claimed to be a better choice in the selection of cluster samples (e.g., Motl et al. 2005; Nagai et al. 2006; Angulo et al. 2009, Planck collab. 2014), we considered 
the sample of clusters 
detected in the survey of the Atacama Cosmology Telescope (ACT) (Hilton et al. 2021).
We matched it to the above eFEDS sample, and only 3 out of 30 clusters did not match in the eFEDS footprint. These clusters are detected with an $S/N<5$, which suggests that the probability of a false detection is 30\% (Hilton et al. 2022).  Moreover, they are not optically identified 
in the deep multiband Hyper Suprime-Cam Subaru survey (Aihara et al. 2022), nor in the Sloan digital sky survey (Ahumada et al 2020).  Our  own visual inspection of the field confirms that there is no galaxy overdensity at the position of the SZ detection, again indicating false detections.
Therefore, these three objects are likely false detections, and the sample is SZ-selected only.

\end{enumerate}
The 
four samples
are considered by the authors above not to be overly affected by cosmic variance.  The three ICM-selected surveys were used to compute luminosities or mass functions based on these data. When limited to $z<0.4$, the comoving volumes sampled by XUCS, ACT, XXL-100, and eFEDS are comparable (the XXL-100 volume is 50\% smaller than that of XUCS, whereas the other surveys cover a volume larger by 50\%). Therefore, comparable numbers of clusters (of any brightness) are expected in these surveys, unless strong selection or evolution  effects are in place. 

\section{Numerical abundance of clusters with a low surface brightness in optical, X-ray and SZ-selected samples}

Fig.~\ref{fig:murcmucod} shows the observed distribution of the optically selected
sample 
in the plane core radius versus mean surface brightness within
a radius of 300 kpc.
The paucity of
$r_c\lesssim150$ kpc at all brightnesses is a planned feature of the XUCS sample, which selects
objects with a velocity dispersion larger than 500 km/s to focus on clusters.
In the XUCS sample, about  one-third of 
the sample have a surface brightness
below 43.35 \subr, 
which is marked in the figure by a blue line. 
This percentage is robust to the precise choice of the minimum core radius for 150 kpc or 100 kpc, for example. We assumed 100 kpc from now on: above this radius, 
about one quarter of the clusters have a low surface brightness.
Their core radiis 
is large, typical of clusters, 
their richness (Puddu \& Andreon 2022) is typical of clusters, but their core-excised luminosity, ($\log L_{500,ce}<43$ \ergs),
is typical of groups.

\begin{figure}
\centerline{\includegraphics[width=9truecm]{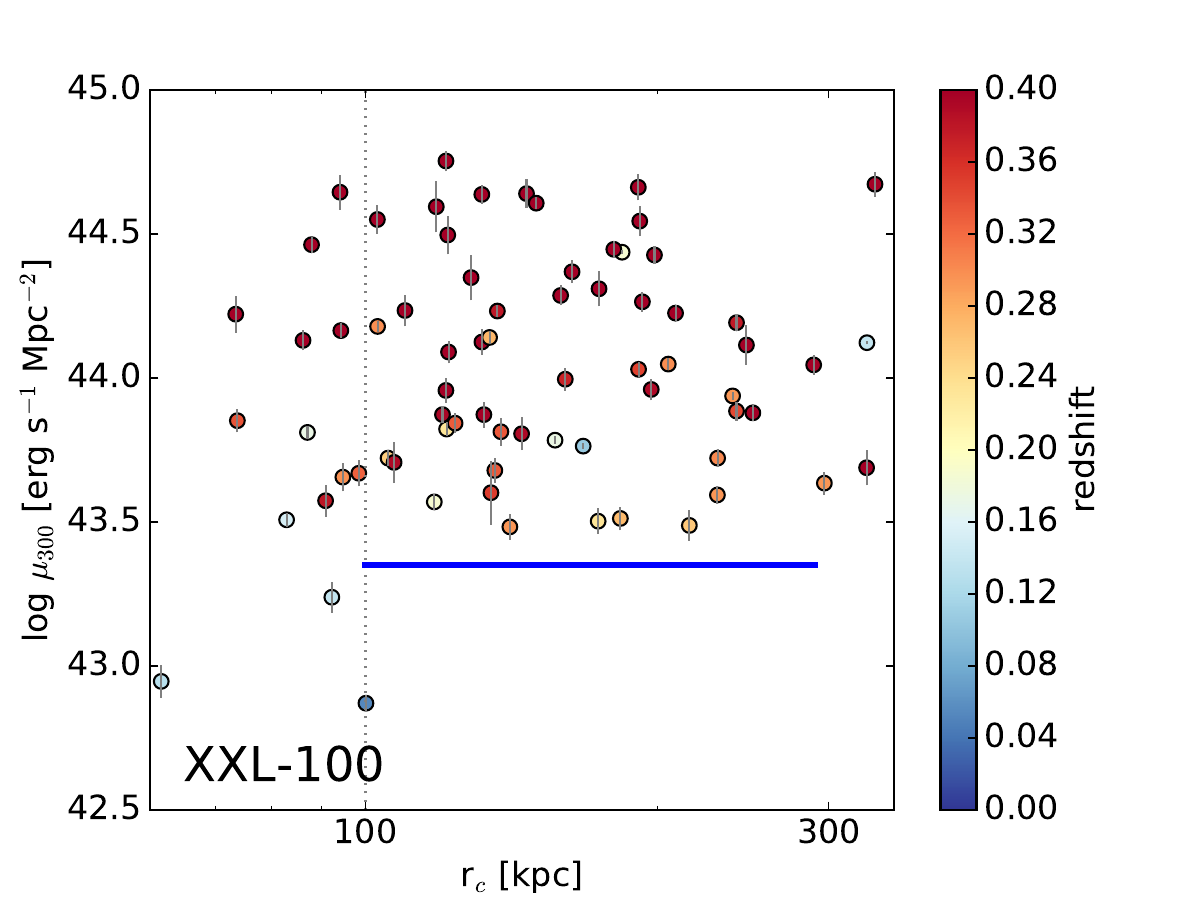}}
\caption[h]{Mean surface brightness within 300 kpc vs core radius for the XXL-100 sample, 
color-coded by redshift (useful for a comparison with Fig.~3 and for the discussion). 
Only about 1\% of the clusters has $r_c>100$ kpc and
a mean surface brightness below 43.35 \subr (blue line) vs one quarter in the XUCS sample.
The vertical dotted line marks the minimum core radius we considered to compute the fraction of clusters with a low surface brightness.
Correcting luminosities for evolution would only alter the points in the top part of the figure,
which is of no interest here. 
}
\label{fig:XXLmurcoffcod}
\end{figure}

Fig.~\ref{fig:XXLmurcoffcod} shows the observed distribution of the XXL-100 sample in the $\mu_{300}-r_c$ 
plane, color-coded by redshift.
One object at  $\mu_{300}<43.35$ \subr among the 56 clusters has $100<r_c<300$ kpc
(i.e., $\sim2$\%)
\footnote{There is one
object of low surface brightness outside the range shown in the Figure, at
$r_c>350$ kpc, 
a region unprobed by XUCS. Visual inspection of the X-ray data 
suggest that it is a bimodal cluster. Even including it, the overall percentage would only be 3\%.}.
Even when we only count objects with $z<0.4$ (and $100<r_c<300$ kpc as before), 
the percentage remains at about 3\%.
The only low surface brightness cluster in XXL-100 has the lowest redshift  in the survey ($z=0.054$, about twice
the distance of the Coma cluster).
This suggests that with the current analysis,
the XXL-100 survey is sensitive enough to reliably detect clusters with a low surface brightness exclusively in the local Universe.
Pacaud et al. (2006) demonstrated that if the clusters had a core radius twice larger than measured, they would largely not be detected in the XXL-100 survey. This agrees with our finding that almost no cluster is found.
An observer willing to study a low X-ray surface brightness with $z>0.054$
in detail would find no cluster in the XXL-100 sample (Fig.~\ref{fig:XXLmurcoffcod}).

The percentage of clusters with a low surface brightness in XXL-100 (between 1\% and 3\%) is significantly lower than the percentage obtained in the XUCS sample, where  about one quarter of the clusters has $r_c> 100$ kpc and $\mu_{300}<43.35$ \subr.
At the opposite end, the XXL-100 sample has a higher percentage of bright 
and compact ($r_c\lesssim 150$ kpc) clusters, which are easy to detect in X-ray selected samples and span a larger volume, as can be seen in the comparison between  
Fig~\ref{fig:XXLmurcoffcod} and Fig~\ref{fig:murcmucod}
at $\mu_{300}\sim 44.5$ \subr .

\begin{figure}
\centerline{\includegraphics[width=9truecm]{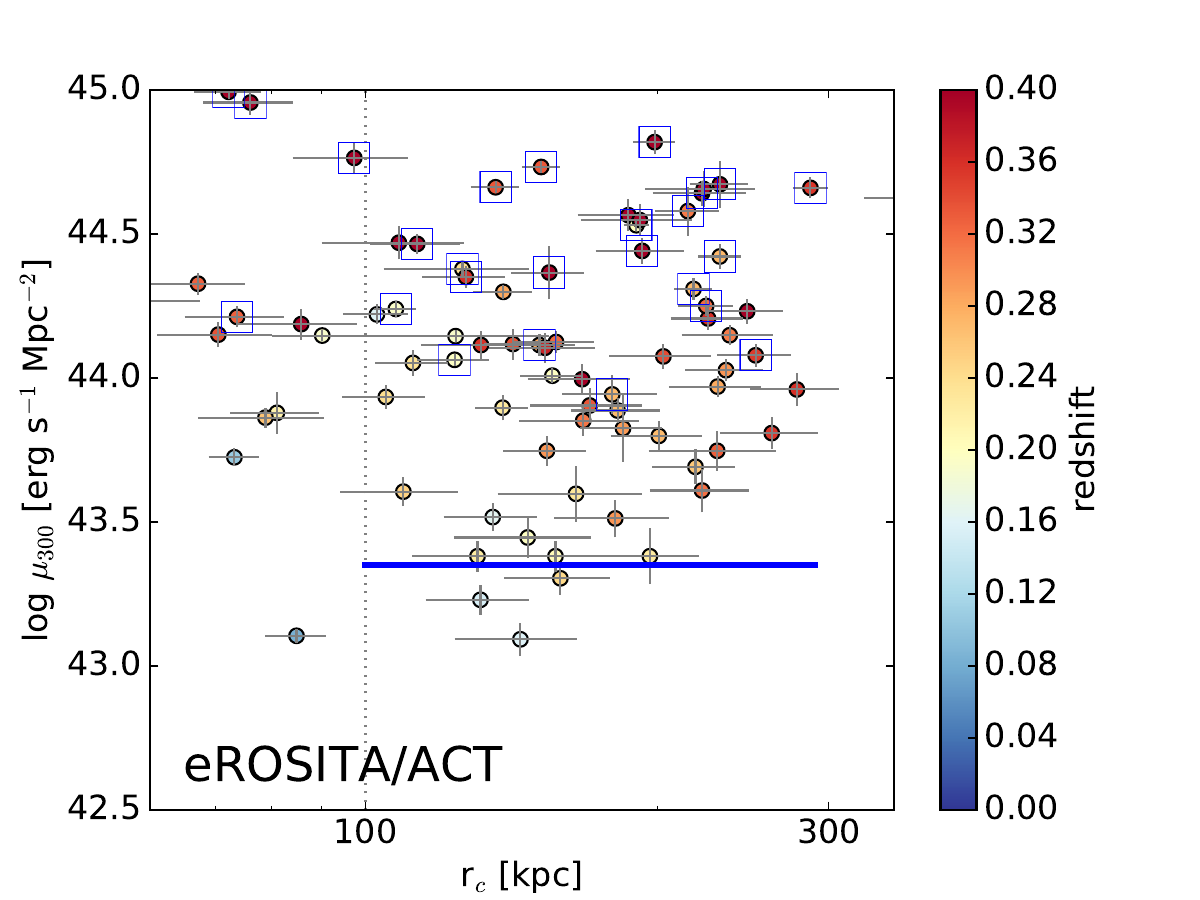}}
\caption[h]{Mean surface brightness within 300 kpc vs core radius,
color-coded by redshift (useful for the comparison with Fig.~2 and for the discussion) for clusters in eFEDS. Clusters that are also detected in the ACT survey are marked with a square. 
About 5\% of the eFEDS clusters have
a mean surface brightness below 43.35 \subr  vs  about one quarter in the XUCS sample.
The ACT survey  only detects the clusters with a higher surface brightness.
The vertical dotted line marks the minimum core radius we considered to compute the fraction of clusters with a low surface brightness. As for the XXL-100 sample,
correcting luminosities for evolution would not significantly alter the results.}
\label{fig:eFEDSmurcoffcod}
\end{figure}

Fig.~\ref{fig:eFEDSmurcoffcod} shows the  observed distribution of the eFEDS sample in the
$\mu_{300}-r_c$ plane, color-coded by redshift.
Only three clusters lie below the blue line and have $100<r_c<300$ kpc. This is 5\% of the sample. The comparison with the  25\% present in the XUCS sample indicates 
that eROSITA  also poorly samples clusters with a low surface brightness. 
As in the case of the XXL-100 sample, eFEDS clusters with a low surface brightness are at low redshift (see the color-coding).
The rarity of clusters with a low X-ray surface brightness agrees with the eFEDS detectability study (Liu et al. 2016)
and with Bulbul et al. (2022).
 
In Fig.~\ref{fig:eFEDSmurcoffcod}
ACT clusters are identified  with a square. 
It is evident from the figure 
that ACT only detects the high surface brightness clusters and does not detect even clusters of
intermediate brightness, $\mu_{300}\sim 43.5-44$ \subr, which are abundant in X-ray selected samples. 
We expect that the South Pole Telescope (SPT) cluster survey (Bleem et al. 2015) 
has the same limitation
since SPT and ACT data share many similarities, such as telescope size and data reduction
(Bleem et al. 2022). 

\begin{figure}
\centerline{\includegraphics[width=9truecm]{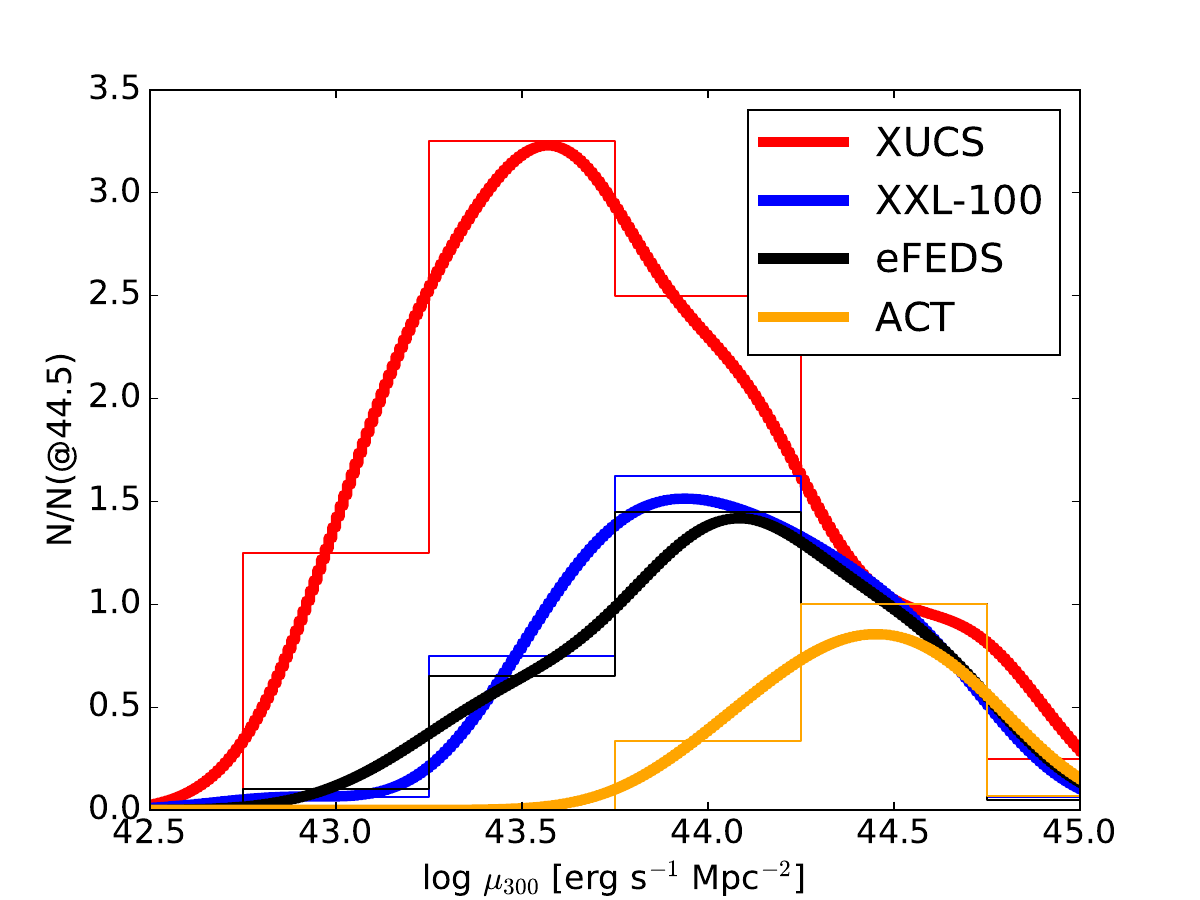}}
\caption[h]{Observed brightness distribution of the various samples normalized at the $\mu_{300} = 44.5$ \subr bin. The histogram
uses top-hat bins, and the curves are running averages using a Gaussian with $\sigma=0.2$ dex. ACT lacks clusters with a
normal brightness ($\mu_{300} \approx 44.0$ \subr), whereas X-ray selected samples  lack most clusters with $\mu_{300} \lesssim 43.7$ \subr.
}
\label{fig:mudistrib}
\end{figure}

So far, we have limited our analysis to the comparison of  the fractions of objects below the arbitrary threshold value of 43.35 \subr .  
We now compare the whole observed surface brightness distributions of the samples (with $100<r_c<300$ kpc as before), although this 
requires that we assume a value for the surface brightness at which the
distributions are normalized. We assumed  
$\mu_{300} = 44.5$ \subr because ACT includes only a few objects at fainter brightnesses, 
and XUCS poorly samples brighter brightnesses.
Fig.~\ref{fig:mudistrib} shows the  observed distribution in brightness for the four samples,
both using top-hat 0.5 dex bins and a running average with 
a Gaussian with $\sigma=0.2$ dex.
The ACT (SZ) sample is shallowest: its brightness distribution only shows rare examples of clusters with an
intermediate brightness of $\mu_{300} \approx 44.0$ \subr  and has no examples with $\mu_{300} \approx 43.5$ \subr or fainter. 
The two X-ray selected sets sample a  wider range than the SZ-selected sample, but they only include a few examples of clusters with $\mu_{300} \sim 43.35$ \subr, 
which are much more abundant in  
XUCS. As mentioned, the normalization is arbitrary and can therefore be changed. However, a normalization change does not displace the location of the peaks of the three distributions horizontally, with XUCS and ACT remaining at the two extremes.  
To summarize, the XUCS sample alone of those considered here provides examples of low surface brightness that are useful for follow-up studies: About one quarter of the sample belongs to this group, as opposed to only a few percent (or none) in the comparison samples.

It might be wondered whether the differences observed in the samples are simply due to differences
in the size of sampled volumes.  As mentioned, XXL-100, eFEDS, and ACT, when bounded to $z<0.4$, sample a comparable comoving volume and should therefore contain a similar number of clusters with $\mu_{300} \sim 43.35$ \subr as XUCS. This is clearly not the case (compare the numbers of points below the blue line in Figs.~\ref{fig:murcmucod}, \ref{fig:XXLmurcoffcod}, and \ref{fig:eFEDSmurcoffcod}), which implies that strong selection effects are at play, given that evolution effects are minor, as discussed in Sec.~4.1. Selection and evolution effects also cause the well-known overabundance of high surface brightness clusters, which are visibile over even larger volumes in ICM-selected samples.

The rarity of clusters with a low surface brightness in X-ray and SZ samples, based on just photometric data (radius and brightness), is derived on purpose without knowledge of the cluster mass or temperature for greater applicability.
Visual inspection of the few objects with a mean surface brightness below 43.35 \subr in the two 
X-ray selected
samples shows that some of them may be groups in which the core radius is temporally inflated by interactions
with a companion (e.g., O'Sullivan et al. 2018), and are not more massive objects in a nearly steady state as seen in XUCS.
A comparison at fixed mass and using a mass-informed aperture, such as
$r_{500}$, would be preferable, but this would 
preclude
the applicability of our work to common cluster samples,
which usually lack a mass estimate.

\section{Discussion}

\begin{figure}
\centerline{\includegraphics[width=9truecm]{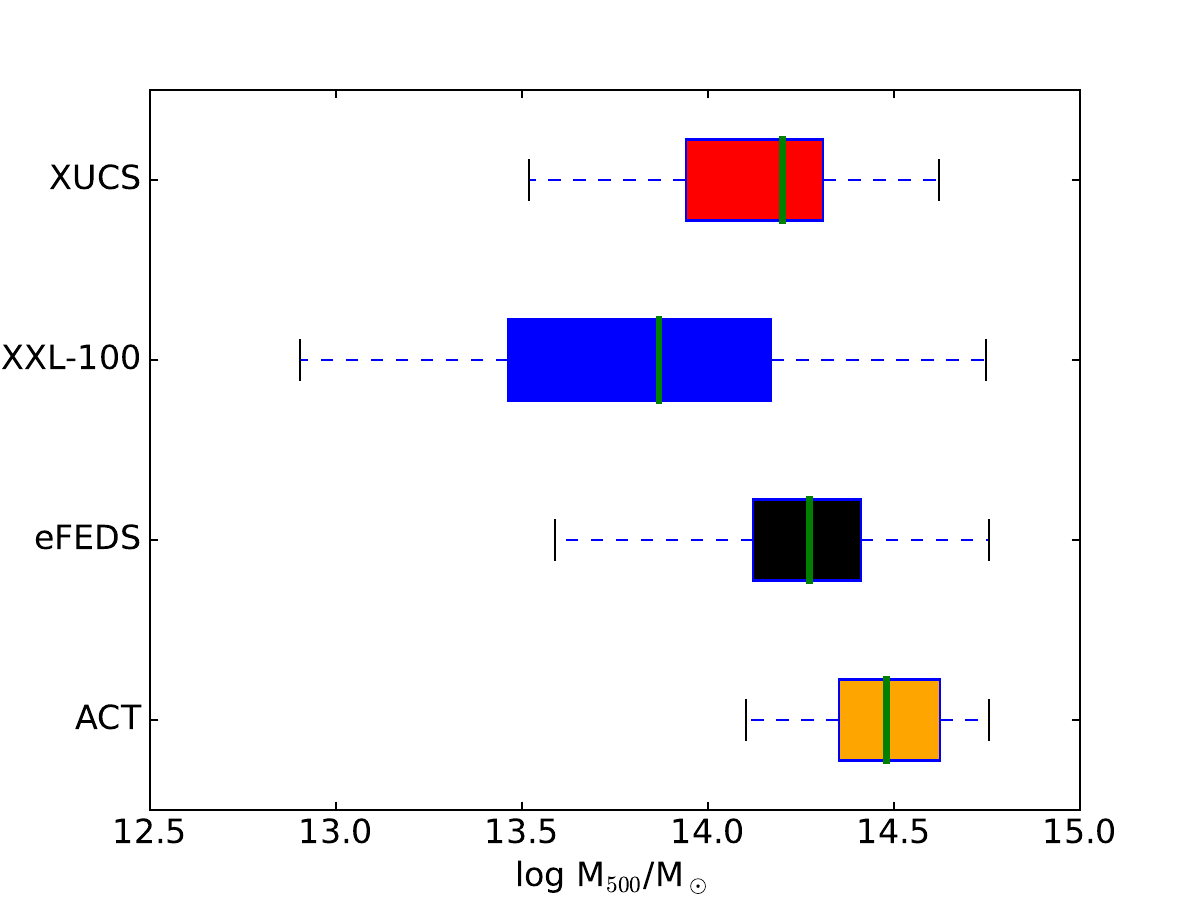}}
\caption[h]{Mass ranges explored by the different cluster samples as a whisker plot. The boxes extend from the lower to the upper quartiles, with a line at the median. The whiskers show the full range.  Mass ranges largely overlap across the samples.
}
\label{fig:massranges}
\end{figure}

In this section, we present our considerations about the results we illustrated above. We note however that they are not exhaustive nor they represents the final word on the subject.

\subsection{Whether clusters of low surface brightness are rare because they are missed, or because they do not exist in the sampled ranges of mass and redshift}

As illustrated above, clusters with a low surface brightness are present in the optically selected sample, but are extremely rare in the ICM-selected samples we considered. 
The question now is whether this arises because we sample a disjoint mass or redshift ranges. 
This is probably not the explanation, as detailed below: the mass and
reshift ranges overlap or are very close to each other.

Fig.~\ref{fig:massranges} shows the mass ranges of the different samples as a whisker plot. Only clusters with $100<r_c<300$ kpc and available masses are considered.
The XUCS sample has caustic masses (Diaferio \& Geller 1997 and later works) with  
$\log M/M_\odot$ values in the range 13.5 to 14.6 (ends of whiskers), and the first, second, and third quartiles (extremes of the box and line inside it) are $\log M/M_\odot=13.90,14.15,$ and $14.30$ (see papers I-IV for details).
Andreon et al. (2017b) found that the caustic masses used in XUCS
are consistent with hydrostatic masses. 
The eFEDS sample includes clusters with mass quartiles similar to those of XUCS (individual mass values are taken from Chiu et al. 2022, 
which are based on count rates that were converted in mass using weak lensing observations).
XXL-100 samples the same mass ranges as XUCS and eFEDS and also includes clusters that are less massive than these samples because it includes
clusters with an estimated weak-lensing mass as low as $\log M/M_\odot=12.7$ (Umetsu et al. 2020). 
The ACT sample is expected to have typical masses higher than $\log M_{500}/M_\odot=14.5$ (Hilton et al. 2021). Their mass range overlaps that of the other samples based on six out of eight clusters with weak-lensing masses in
Miyatake et al. (2019). According to the values tabulated in Chiu et al. (2022) and shown in Fig.~\ref{fig:massranges}, most of the sample is in the top three quartiles for XUCS and eFEDS.

With XUCS, we have shown 
that the Universe at $0.050<z<0.135$, which is included in the redshift range probed by both XXL-100 and eFEDs, 
contains clusters with a low surface brightness whose masses are sampled by both XXL-100 and eFEDs (see also other detections by Lietzen et al. 2024; O'Sullivan et
al. 2017; Pearson et al. 2017; Xu et al. 2018). Therefore, clusters with a low surface brightness are known to exist in the same range of
redshift and mass as  is probed by these two surveys, but they are rare at best in these catalogs. XXL-100 reaches lower masses than XUCS (Fig.~\ref{fig:massranges}).

Since the redshift ranges of the above samples overlap, it is unlikely that evolution plays a large role in explaining the different faint ends of the brightness distribution.
Furthermore,
according to the determinations of Giles et al. (2016) and Chiu et al. (2022) for the XXL-100 and eFEDS samples, the evolution is minimum at low redshift,
so that we expect that they are present in the whole wider redshift range probed by these surveys.
To match the peaks of X-ray selected and XUCS samples, a 0.5 dex evolution is
needed on average (Fig.~\ref{fig:mudistrib}) in the 0.5 Gyr just before $z=0.08$,
which is huge and unlikely to have been so badly mistaken by these authors. 
Finally, the faint end of the brightness distribution of these two surveys cannot change because in their samples the objects with a lower surface brightness are at low redshift and therefore have negligible evolutionary corrections. 
Therefore, low surface brightness clusters 
exist 
in the redshift and mass ranges explored by these X-ray surveys, 
and their rarity is intrinsic in these catalogs. 

The case for the ACT survey is less clear. The survey is mostly sensitive to clusters
with expected masses $\log M_{500}>14.5$, according to Hilton et al. (2021). However, according to Miyatake et al. (2019) and Chiu et al. (2022), 
the ACT mass range largely overlaps the range explored by the other samples (Fig.~\ref{fig:massranges}). The ACT clusters considered in this work have $z>0.19$, which is outside the range probed by XUCS.
The difference in redshift probed is not enough to invoke evolution to explain the difference unless a) an extreme evolution is assumed, and b)
 both Giles et al. (2016) and Chiu et al. (2022) are in error.
Therefore, 
we currently cannot firmly conclude that
ACT does not contain examples of clusters with a low surface brightness  because they do not exist at the mass/z of the survey.
However, at intermediate brightness (e.g., 43.5-44.3 \subr), 
eFEDS contains clusters at $z\sim0.3$ with large $r_c$, but only some of them are detected by ACT (the orange-red points in the center right panel of Fig.~\ref{fig:eFEDSmurcoffcod}). This indicates that intermediate-brightness clusters in the region of the Universe that is covered by ACT exist, but they are not detected.  
Further support for the existence of massive clusters that are undetected by ACT comes from a) the Orlowski-Scherer et al. (2021)
analysis, which required half of the richest clusters to be missed by ACT; b) the very rich but SZ-faint clusters in Dicker et al. (2020) and in Di Mascolo et al. (2020); and c) the low SZ strength signal for its mass of JKCS\,041, a $z=1.803$ cluster (Andreon et al. 2023). The redshift of all these
clusters is much higher than probed in this work, but they indicate that massive clusters with a low SZ signal also exist outside of the very local Universe.

To summarize, the two X-ray surveys do not include existing low surface brightness clusters, and the ACT survey does not include existing intermediate surface brightness
clusters. This suggests that the  paucity of low to intermediate surface brightness clusters in the X-ray and SZ catalogs is not due to the fact that these clusters do not exist in the mass/z range explored
by these catalogs, but that they have been missed at the current sensitivity of the surveys.

The authors of X-ray and SZ surveys are aware that clusters with a low brightness are difficult to detect in X-ray surveys
(Vikhlinin et al. 2008; Moretti et al. 2004; Pacaud et al. 2006, Pacaud et al. 2016, see Andreon et al. 2019 and 2022 for a detailed listing).
Figs.~2
and 3 in Andreon et al. (2022) showed that it is challenging to account for low surface brightness clusters for the $L_X-M$ and $T-M$ scaling relations because 
a smaller scatter around the mean relation is derived, even when the X-ray selection is said to be accounted
for, compared to the mean relation observed in samples that include the low surface brightness population. As detailed in Andreon et al. (2019) and as was at least known since Vikhlinin et al. (2009b), the determination of the scaling relation parameters is conditional on assumptions on the unseen population, namely being able to estimate how many low surface brightness clusters are missed, and how these objects populate the plane of the scaling relation in question. These determinations are extremely hard to make when the collected sample of low surface brightness objects is a few clusters or none. Our work shows that low surface brightness clusters exist.
Their impact on the scaling relations
may range from negligible (for the $L_X-T$ scaling relation; Andreon et al. 2022) to major (for the $T-M$ and $L_X-M$ relations; Andreon et al. 2016, 2022) and
depends on survey characteristics.

\begin{figure}
\centerline{\includegraphics[width=9truecm]{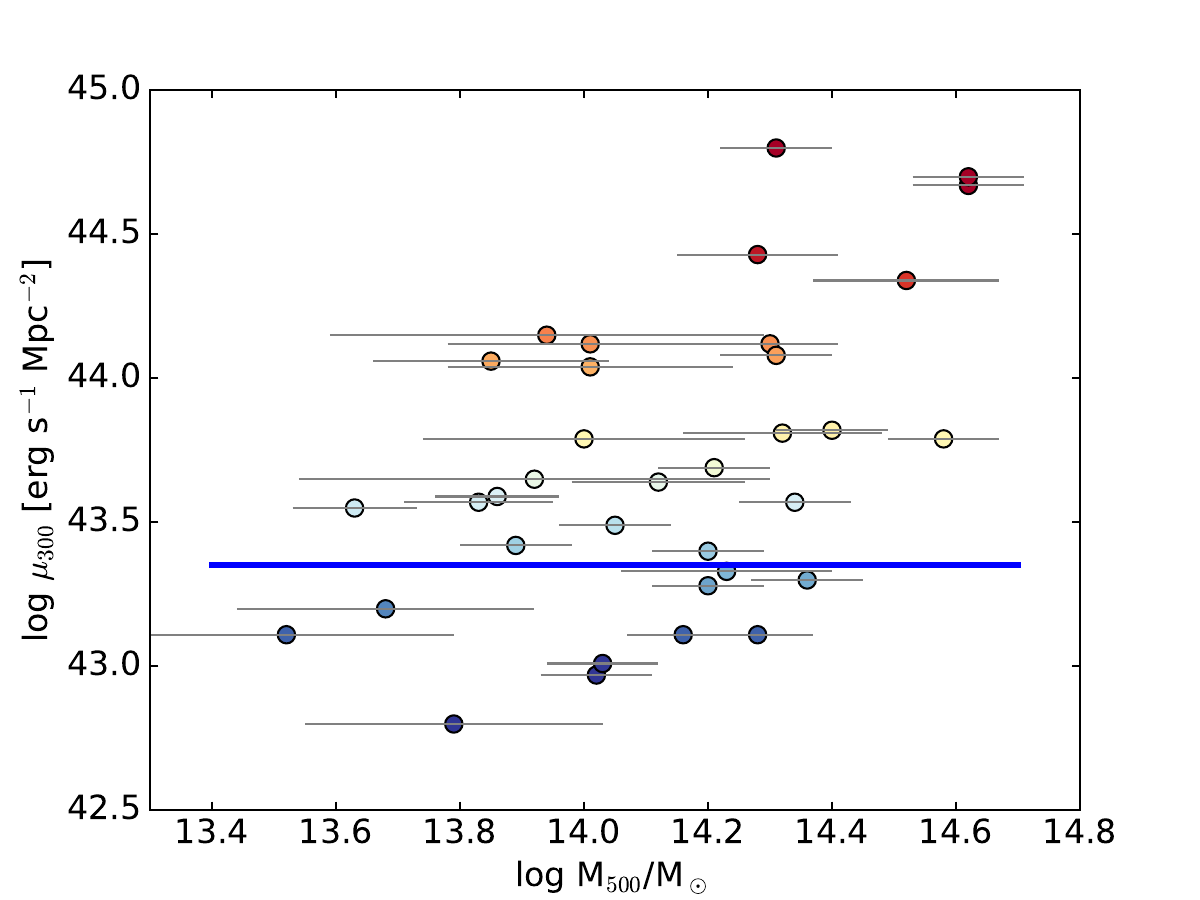}}
\caption[h]{Mean surface brightness $\mu_{300}$ in the [0.5-2] keV band vs $M_{500}$ for the XUCS sample.
Clusters with a low surface brightness are spread over three out of the four mass quartiles. The large scatter between
these two quantities strongly reduces the impact of the differences in mass on the brightness distribution.
We plot the measurements of $\mu_{300}$ from
two different telescopes for three clusters. 
}
\label{fig:mu300Moffcod}
\end{figure}

\begin{figure}
\centerline{\includegraphics[width=9truecm]{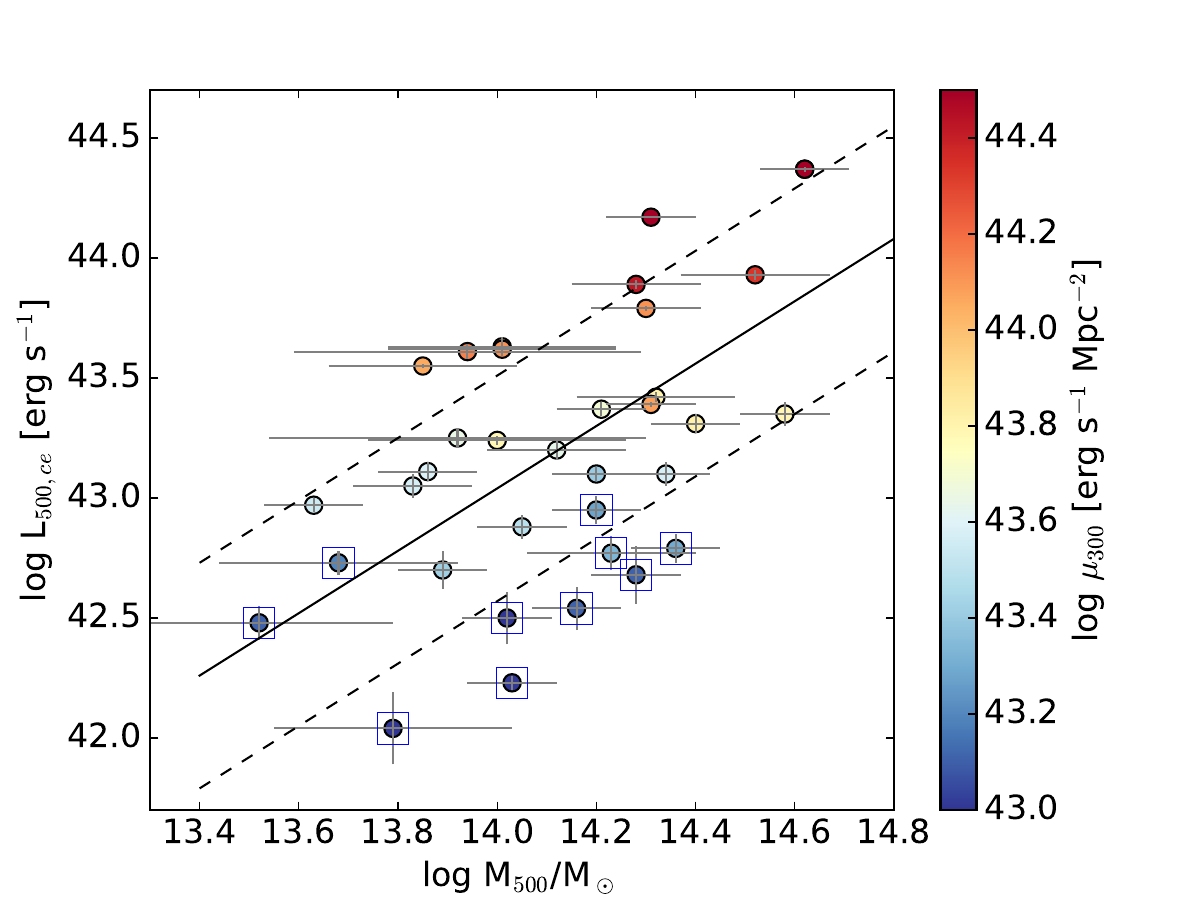}}
\centerline{\includegraphics[width=9truecm]{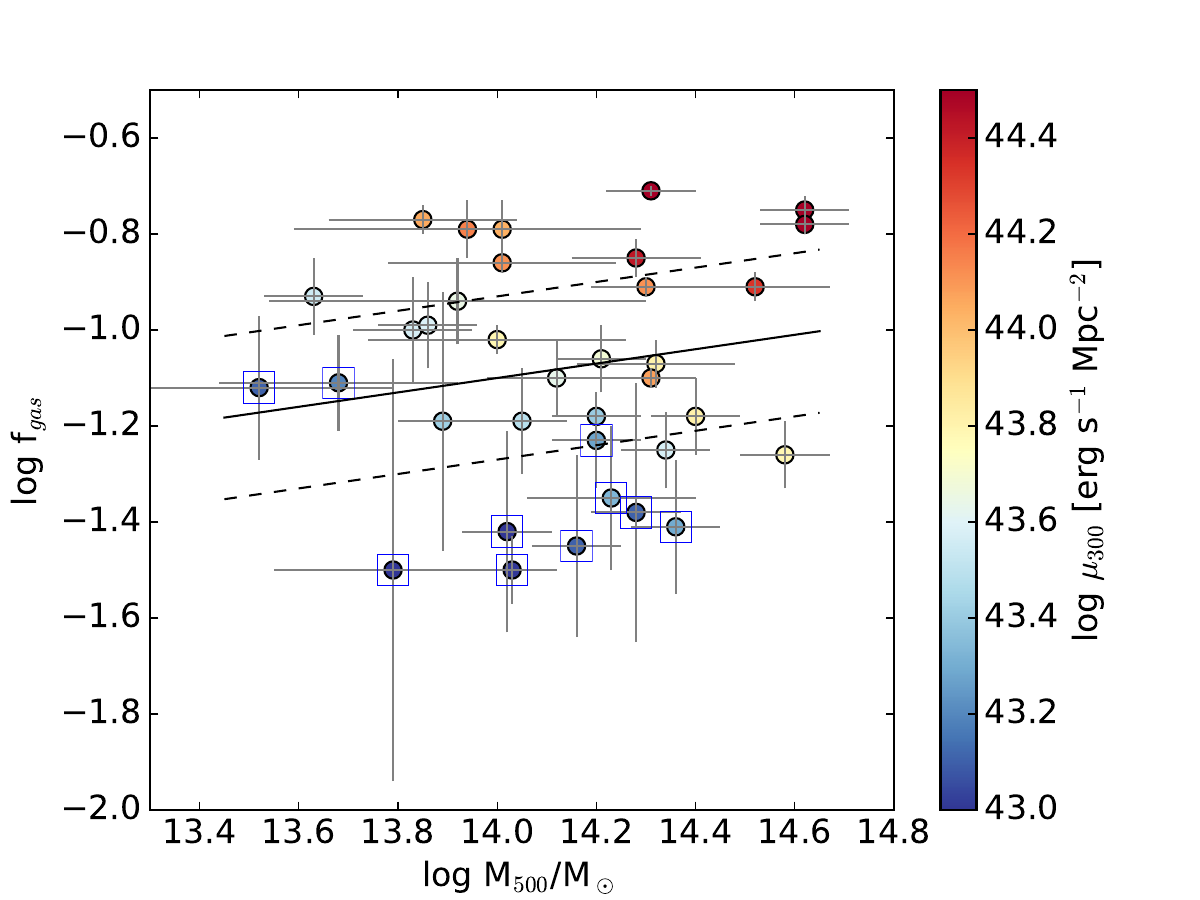}}
\caption[h]{Core-excised [0.5-2] keV luminosity (upper panel) and gas fraction (bottom panel)
within $r_{500}$ vs $M_{500}$ color-coded by brightness within 300 kpc
for the XUCS sample. 
The solid black line and the dashed corridor indicate the mean relation and
the $\pm 1\sigma_\textrm{intr}$ regions derived in Papers I and II. 
Low surface brightness clusters (with $\mu_{300}<43.35$ \subr) are indicated by squares and are mostly 
clusters that are X-ray faint for their mass and have a low gas fraction (near or below the lower dashed
line)
with some contamination by objects with an average gas fraction that are faint because the mass is low (near the
solid line).
We plot the measurements from
two different telescopes for three telescopes, although it can be challenging to identify these duplicates in the plot. 
}
\label{fig:LxMoffcod}
\end{figure}

\subsection{Whether the comparison between samples is influenced by a tight brightness-mass
correlation and by large mass biases}

If very hypothetically the masses of XUCS clusters were strongly overestimated and if there were a tight and steep brightness-mass relation, then the XUCS
brightness distribution would include clusters with a low brightness that would be rightfully absent in the X-ray selected samples. We now show that neither of these
two hypotheses, which are both needed to bias our results, is true. The XUCS masses are not systematically overestimated (Papers I, III, and IV), and a huge mass overestimate is
needed to make them less massive than eFEDS and XXL-100 clusters with which we compare them. For example, the most massive low surface
brightness cluster, CL2015, has a caustic mass of $\log M_{500}=14.39\pm0.09$ (Paper I) and a hydrostatic mass of $\log M_{500}=14.23\pm0.22$ (Paper IV).
A mass bias of 0.6 dex by both the caustic and hydrostatic method, which is unheard of before, is needed to bring the
CL2015 mass below the range sampled by XXL-100. Second, the relation between brightness and
mass is neither tight nor steep: Fig.~\ref{fig:mu300Moffcod} shows the distribution of the XUCS sample in the mean surface brightness $\mu_{300}$ versus mass $M_{500}$ plane, with mass errors including
those induced by triaxiality and projection effects. Because of the large scatter, a hypothetical over- or underestimate of the mass has a negligible effect on the XUCS brightness
distribution (the effect would be exactly zero in absence of a relation) and on the results of our comparison of the samples.

\subsection{Robustness analysis}

Although the details of the radial profile fit are not the same, the core radii in XUCS, XXL-100, and eFEDS are all derived from a fit to the X-ray photons with a circular beta model (possibly with a cusp in the case of XUCS) with the same fixed beta parameter and allowing for a background. We only used the core radii to remove very compact objects with sizes typical of groups (smaller than 100 kpc) from the sample.  Moreover, the fraction of low surface brightness clusters does not depend strongly on the core radius (see Figs~\ref{fig:murcmucod} to ~\ref{fig:eFEDSmurcoffcod}), and therefore,  we do not expect that the slightly different methods for deriving the core radius have any significant effect on the final results.

As in Andreon et al. (2022), 2 out of 34 clusters  were excluded from the XUCS sample.
Returning them (at whatever surface brightness) in XUCS
would not cause low surface brightness clusters to appear in other samples, such as eFEDS, XXL-100, or ACT samples.
Independently of whether these
two clusters have a low or high surface brightness, the fraction of low surface brightness clusters of XUCS continues to be about one quarter, and our conclusions about the rarity of clusters of low surface brightness in X-ray and SZ- selected samples are not altered. Regardless of the surface brightness of the two missing XUCS clusters, the ACT sample lacks intermediate-brightness clusters compared to the X-ray selected samples.

\subsection{A first look at the ICM properties of clusters that are rare in ICM-selected samples}

Fig.~\ref{fig:LxMoffcod}  shows 
the XUCS $L_X-M$ and $f_{gas}-M$ relations that were presented in previous papers, but
with the points color-coded by surface brightness.  Clusters with 
$\mu_{300}<43.35$ \subr, which we have shown to be rare
in ICM-selected samples, 
are indicated by squares in the figure. 
Clusters with a low X-ray luminosity for their mass and with a low gas fraction lie well below the mean line (near or below the lower dashed line). All of them have 
a low surface brightness.
However, the low surface brightness clusters (i.e., squares)
include clusters near the solid line that are
faint because they have a low mass and a common gas fraction. 
To identify clusters with a low luminosity for their mass (and with low gas fraction) among those with a low surface brightness,
a mass determination is needed.
Because of the large scatter in the $M-T$ relation (Andreon et al. 2022), the
mass cannot be inferred from $T$
following the common practice of estimating $r_{500}$ from $T$.
Similarly, 
$f_{gas}$, $M_{gas}$, or $Y_X$  cannot be used for this purpose (see Andreon et al. 2022 for details). To obtain a reliable estimate of the total mass, 
we would need a measure that samples the regions 
that contain most of the mass, such as a hydrostatic estimate with a precisely measured
temperature gradient near $r_{500}$, or weak-lensing or caustic masses, which all are costly observationally.
A preselection on $\mu_{300}$ would therefore significantly reduce the number of targets to be followed-up for mass determination.

\section{Conclusions}

The comparison of the properties of galaxy cluster samples selected using observations in different wavebands
may shed light on potential biases of the way in which the samples are assembled. For this comparison, we introduced a new observable 
that does not require previous knowledge of the cluster mass: The X-ray
mean surface brightness within the central 300 kpc.

We found that clusters with a low surface brightness, defined as those with a mean surface brightness below 43.35 \subr within 300 kpc in the [0.5-2] keV band, are  about one quarter of the whole cluster population in a sample of 32 clusters in the nearby Universe selected independently of the
intracluster medium properties. On the other hand, almost no example of a low central surface brightness cluster exists in
two X-ray selected samples, one based on XMM-Newton XXL-100 survey data and one using full-depth eROSITA eFEDS data, even though they are known to exist in the same range of redshift and mass as probed by these two surveys.

Furthermore, 
the
Sunayev-Zeldovich Atacama Cosmology Telescope cluster survey
is even more selective than the previous two samples because it  does not even include clusters with an
intermediate surface brightness, which are instead present in X-ray selected samples that explore the same volume of the Universe.

Finally, a measure of the
mean surface brightness, which is obtained without knowledge of the mass, 
proves to be effective in
narrowing the number of clusters to be followed-up to recognize those
with a low gas fraction or with a low X-ray luminosity for their mass,
whose identification would otherwise require knowledge of the mass for all clusters. 

Since the introduced observable does not require previous knowledge of the cluster mass, it is possible and
useful to repeat the work performed here using other cluster samples and to extend it considering a
sample with tighter redshift overlaps, for example.

\begin{acknowledgements}
We thank Dominique Eckert and an anonymous reader for constructive discussion.
S.A. acknowledges financial contribution from the agreement ASI-INAF n.2017-14-H.0 
This work has been partially supported by the ASI-INAF program I/004/11/5.
\end{acknowledgements}

{}

\appendix{}

\section{List of core radii and brightnesses}
\begin{table*}
\caption{Results of the analysis.}
\begin{tabular}{lrrrrrrr}
\hline\hline
Id  & \multicolumn{1}{c}{RA} & Dec & \multicolumn{1}{c}{z} & $r_c$ & err & $\log \mu_{300}$ & err \\
   & \multicolumn{2}{c}{(J2000)}&  & \multicolumn{2}{c}{kpc}  & \subr & dex \\
\hline
 CL1001 & 208.2560 &  5.1340  & 0.079   & 240 & 12 & 44.12 & 0.01 \\ 
 CL1009 & 198.0567 & -0.9744  & 0.085   & 149 & 8 & 43.79 & 0.01 \\  
 CL1011 & 227.1073 & -0.2663  & 0.091   & 167 & 26 & 43.33 & 0.04 \\ 
 CL1014 & 175.2992 &  5.7350  & 0.098   & 180 & 25 & 43.82 & 0.02 \\ 
 CL1015 & 182.5701 &  5.3860  & 0.077   & 183 & 11 & 44.06 & 0.01 \\ 
 CL1018 & 214.3980 &  2.0530  & 0.054   & 178 & 22 & 43.20 & 0.04 \\ 
 CL1020 & 176.0284 &  5.7980  & 0.103   & 159 & 16 & 44.15 & 0.02 \\ 
 CL1030 & 206.1648 &  2.8600  & 0.078   & 164 & 35 & 43.11 & 0.05 \\ 
 CL1033 & 167.7473 &  1.1280  & 0.097   & 142 & 21 & 43.57 & 0.04 \\ 
 CL1038 & 179.3788 &  5.0980  & 0.076   & 239 & 25 & 43.69 & 0.03 \\ 
 CL1039 & 228.8088 &  4.3860  & 0.098   & 193 & 15 & 43.81 & 0.01 \\ 
 CL1041 & 194.6729 & -1.7610  & 0.084   & 153 & 2 & 44.80 & 0.01 \\  
 CL1047 & 229.1838 & -0.9693  & 0.118   & 212 & 26 & 44.04 & 0.03 \\ 
        &          &          &         & 297 & 21 & 44.12 & 0.01 \\ 
 CL1052 & 195.7191 & -2.5160  & 0.083   & 140 & 3 & 44.08 & 0.01 \\  
 CL1067 & 212.0220 &  5.4180  & 0.088   & 220 & 25 & 43.65 & 0.03 \\ 
 CL1073 & 170.7265 &  1.1140  & 0.074   & 269 & 22 & 43.40 & 0.02 \\ 
        &          &          &         & 212 & 38 & 43.28 & 0.04 \\ 
 CL1120 & 188.6107 &  4.0560  & 0.085   & 153 & 27 & 42.97 & 0.08 \\ 
 CL1132 & 195.1427 & -2.1340  & 0.085   & 92 & 29 & 42.80 & 0.09 \\  
 CL1209 & 149.1609 & -0.3580  & 0.087   & 87 & 7 & 43.01 & 0.02 \\   
 CL2007 & 46.5723  & -0.1400  & 0.109   & 209 & 38 & 43.11 & 0.09 \\ 
 CL2010 & 29.0706  &  1.0510  & 0.080   & 222 & 27 & 43.57 & 0.03 \\ 
 CL2015 & 13.9663  & -9.9860  & 0.055   & 184 & 25 & 43.30 & 0.03 \\ 
 CL2045 & 22.8872  &  0.5560  & 0.079   & 190 & 22 & 43.59 & 0.03 \\ 
 CL3000 & 163.4024 & 54.8700  & 0.072   & 200 & 27 & 43.79 & 0.02 \\  
 CL3009 & 136.9768 & 52.7900  & 0.099   & 119 & 25 & 43.42 & 0.05 \\ 
 CL3013 & 173.3113 & 66.3800  & 0.115   & 167 & 12 & 44.43 & 0.01 \\ 
 CL3020 & 232.3110 & 52.8600  & 0.073   & 195 & 26 & 43.64 & 0.03 \\ 
 CL3023 & 122.5355 & 35.2800  & 0.084   & 144 & 22 & 43.49 & 0.04 \\ 
 CL3030 & 126.3710 & 47.1300  & 0.127   & 233 & 17 & 44.34 & 0.01 \\ 
 CL3046 & 164.5986 & 56.7900  & 0.135   & 257 & 15 & 44.67 & 0.01 \\ 
        &          &          &         & 261 & 9 & 44.70 & 0.01 \\  
 CL3049 & 203.2638 & 60.1200  & 0.072   & 123 & 14 & 43.55 & 0.02 \\ 
 CL3053 & 160.2543 & 58.2900  & 0.073   & 105 & 21 & 43.11 & 0.09 \\ 
\hline
\end{tabular}
\hfill \break 
Three clusters were observed by two telescopes and therefore are listed twice.
Id, RA, Dec, and $z$ are from Paper I. \hfill
\label{tab1}
\end{table*}

\label{lastpage}

\end{document}